\RequirePackage{lineno}
\documentclass[twocolumn,aps,prd,groupedaddress,nofootinbib,showpacs]{revtex4-1}
\usepackage{graphicx}
\usepackage{lipsum}
\usepackage{amsmath}
\usepackage{bm}
\usepackage{slashed}
\usepackage{epsfig}
\usepackage{epstopdf}
\usepackage{amsfonts}
\usepackage{subfigure}
\usepackage{color}
\usepackage{diagbox}
\usepackage[pdftex]{hyperref}
\newcommand{\sect}[1]{\section{#1}}
\begin{document}

\title{Improvement for Color Glass Condensate factorization: single hadron production in pA collisions at next-to-leading order}

\author{Hao-Yu \surname{Liu}}
\affiliation{School of Physics and State Key Laboratory of Nuclear Physics and Technology, Peking University, Beijing 100871, China}
\author{Yan-Qing \surname{Ma}}
\affiliation{School of Physics and State Key Laboratory of Nuclear Physics and Technology, Peking University, Beijing 100871, China}
\affiliation{Center for High Energy Physics, Peking University, Beijing 100871, China}
\affiliation{Collaborative Innovation Center of Quantum Matter, Beijing 100871, China}
\author{Kuang-Ta \surname{Chao}}
\affiliation{School of Physics and State Key Laboratory of Nuclear Physics and Technology, Peking University, Beijing 100871, China}
\affiliation{Center for High Energy Physics, Peking University, Beijing 100871, China}
\affiliation{Collaborative Innovation Center of Quantum Matter, Beijing 100871, China}

\date{\today}

\begin{abstract}
High order calculation at semi-hard scale in high energy collisions is very important, but a satisfactory calculation framework is still missing. We propose a systematic method to regularize the rapidity divergence in CGC factorization, which makes higher order calculation rigorous and straight forward. By applying this method to the single hadron production in the pA collision, we find the kinematic constraint effect introduced by hand in previous works comes out automatically, but with different values. The difference is crucial for our next-to-leading order (NLO) result to have a smaller theoretical uncertainty compared with LO result, which makes the high order calculation in CGC factorization to be useful. As a byproduct, the negativity problem found in the literature can also be overcome in our framework by properly choosing the factorization scale. 
\end{abstract}

\maketitle

\allowdisplaybreaks

\sect{Introduction}
In high energy collision with small Bjorken $x$, the nonlinear effect of QCD will compete with linear BFKL evolution effect \cite{Balitsky:1978ic,Lipatov:1976zz,Kuraev:1976ge,Fadin:1975cb}, which leads to the gluon saturation phenomenon \cite{Gribov:1984tu,Mueller:1985wy}. The saturation scale $Q_s(x)$, characterizing the typical transverse momentum of gluons with momentum fraction $x$, gets larger for smaller $x$. When the semi-hard scale $Q_s(x)$ is large enough, asymptotic freedom of QCD enables us to apply perturbative QCD to study the small-$x$ gluon distributions. However, large-$x$ gluons always present in any physical process, which makes it hard to take advantage of perturbation theory.

The Color Glass Condensate (CGC) \cite{Iancu:2002xk,Iancu:2003xm,Gelis:2010nm,Kovchegov:2012mbw} is an effective field theory of QCD that can distinguish large-$x$ gluons from small-$x$ gluons in a target. Large-$x$ gluons, with very long light-cone lifetime, are not created or annihilated during collision, and they serve as a background color source for small-$x$ gluons. The large-$x$ gluons are stochastically distributed in each collision event, and a CGC average of large-$x$ distributions is allowed to define them as universal nonperturbative quantities. Although small-$x$ gluons can be created and annihilated during collision, their effects can be calculated perturbatively. For quantitative purpose, a parameter $X_f$ is introduced and define small $x$ by $x<X_f$. The evolution of CGC-averaged nonperturbative quantities w.r.t. $X_f$ is governed by the renormalization group equation, called JIMWLK equation\cite{JalilianMarian:1997jx,JalilianMarian:1997gr,Kovner:2000pt,Iancu:2000hn,Iancu:2001ad,Ferreiro:2001qy}, which should cancel exactly with the $X_f$-dependence of small-$x$ effects, and thus leaves physical observables independent of the separation value $X_f$.  Things are significantly simplified in the large $N_c$ limit, where many processes are found to only depend on color dipole scattering amplitudes, which satisfy a closed Balitsky-Kovchegov (BK) evolution equation\cite{Balitsky:1995ub,Kovchegov:1999yj}. With the JIMWLK or BK equation, the only 
unknown information in the CGC effective theory is the CGC-averaged nonperturbative quantities at a given separation value $X_0$, which can be modeled with some parameters to be determined by experimental data \cite{McLerran:1993ni,McLerran:1993ka}.

Because $\alpha_s$ is not so small at semi-hard region, a LO calculation in CGC factorization is usually not enough. High order calculations are very crucial for theoretical predictions to have sufficient accuracy.

In the following, we will concentrate on single inclusive hadron production at forward rapidity region in nucleon-nucleus (pA) collision \cite{Dumitru:2002qt,Dumitru:2005gt,Kovchegov:1998bi,Kovchegov:2001sc,Albacete:2003iq,Kharzeev:2003wz,Iancu:2004bx,Blaizot:2004wu,Blaizot:2004wv,Albacete:2010bs,Tribedy:2011aa,Tribedy:2011aa,Lappi:2013zma}. The first NLO calculation was obtained in Refs.~\cite{Chirilli:2011km,Chirilli:2012jd}, which however results in negative cross sections for high transverse momentum $p_{h\perp}$ hadron production \cite{Stasto:2013cha}. 
Several works attempt to solve the problem \cite{Kang:2014lha,Altinoluk:2014eka,Watanabe:2015tja,Ducloue:2016shw,Iancu:2016vyg}, which either gives up standard factorization structure or still suffers from negativity problem. Especially, it was found that a kinematic constraint needs to be introduced by hand based on physical arguments \cite{Watanabe:2015tja,Iancu:2016vyg}. It tells us that a satisfactory framework for CGC factorization at high order is still missing.

In this paper, we propose a systematic method to do high order calculation in CGC factorization. Like any standard factorization theory, theoretical uncertainties caused by the missing of higher order contributions can be estimated by varying the factorization scale $X_f$. The effect of kinematic constraint appears automatically in our method, but with a different value. It is the difference that makes our NLO result to have a smaller theoretical uncertainty comparing with LO result. As a byproduct, the negativity problem can be overcome by a proper choosing of factorization scale.

\sect{Kinematics and leading order result}
We use collinear factorization to deal with the initial-state proton and the final-state observed hadron. As for the interaction between the probe parton and the target, it is proper to apply CGC effective theory. Then the differential cross section can be formally factorized as $d\sigma=f(x)\otimes D(z){\otimes}\mathcal{H}$, where $f(x)$ stands for PDF, $D(z)$ stands for fragmentation function (FF), and $\mathcal{H}$ includes both impact factors and nonperturbative functions defined in CGC. To calculate $\mathcal{H}$, it is convenient to use light-cone perturbation theory \cite{Brodsky:1997de,Lepage:1980fj} in momentum space. 

\begin{figure}
	\centering
	\subfigure[]{
		\label{fig:lo}
		\includegraphics*[scale=0.4]{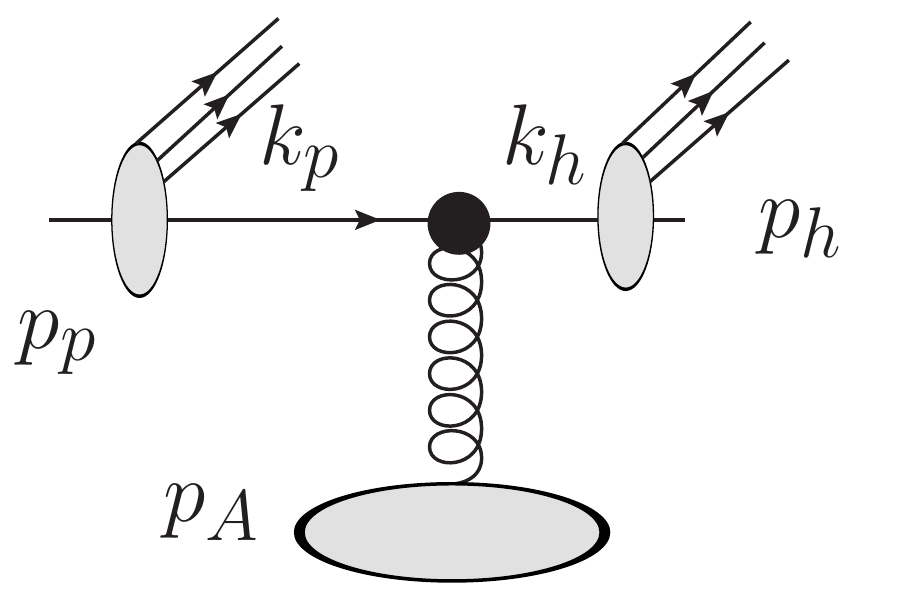}}
	\subfigure[]{
		\label{fig:vc.1}
		\includegraphics*[scale=0.4]{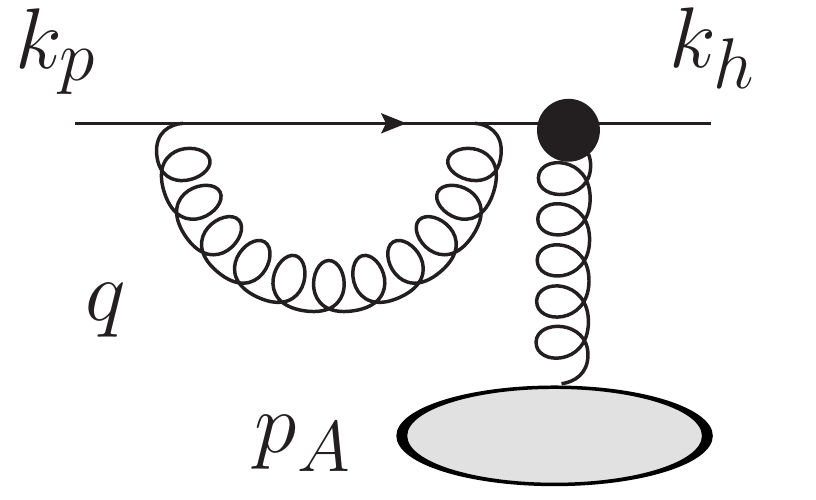}}
	\subfigure[]{
		\label{fig:vc.2}
		\includegraphics*[scale=0.4]{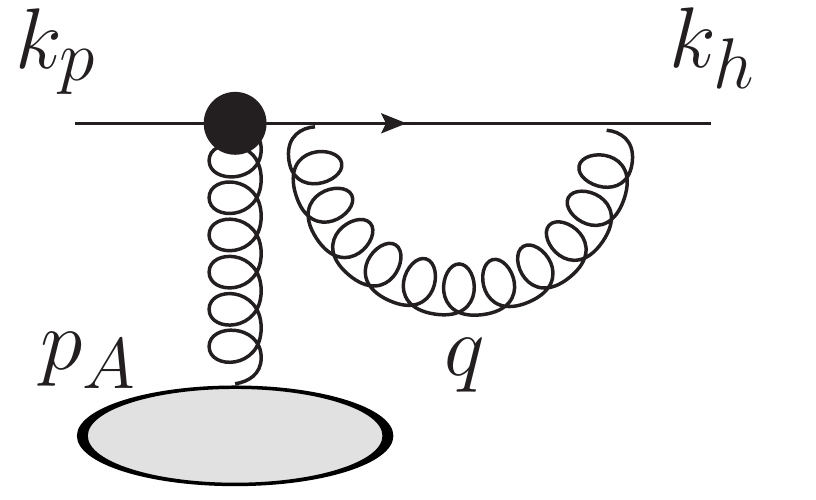}}
	\subfigure[]{
		\label{fig:vc.3}
		\includegraphics*[scale=0.4]{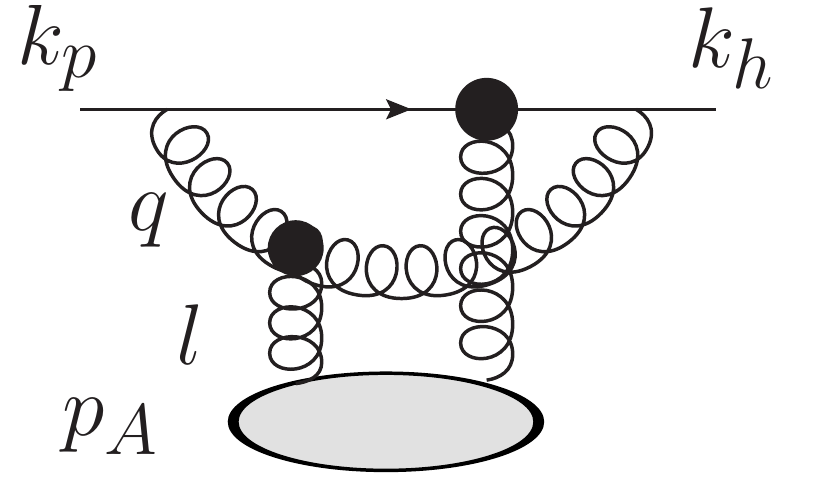}}
	\subfigure[]{
		\label{fig:rc.1}
		\includegraphics*[scale=0.4]{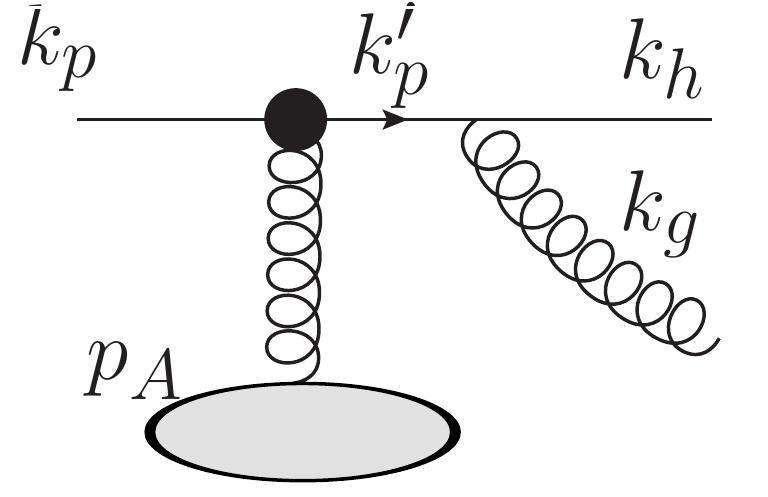}}
	\subfigure[]{
		\label{fig:rc.2}
		\includegraphics*[scale=0.4]{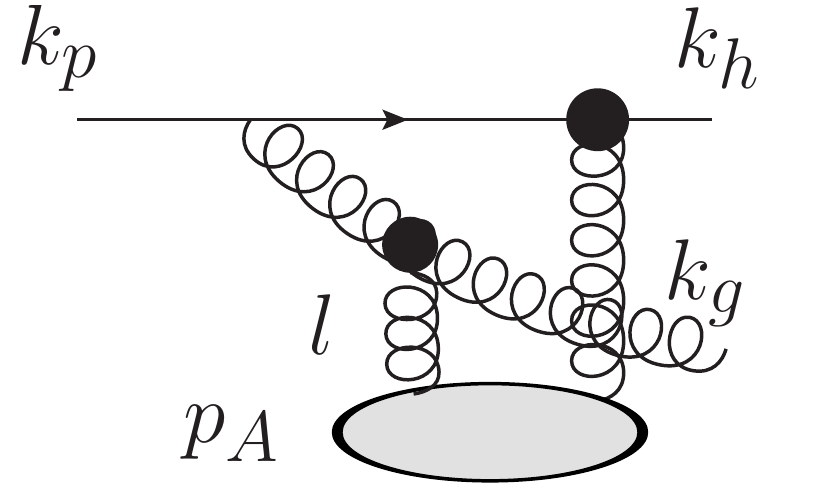}}
		\caption{\label{fig:diag} Feynman diagrams at the LO and the NLO for $q+A\to q+X$ process. (a) is the LO Feynman diagram, (b)-(d) are diagrams for the virtual correction, and (e) and (f) are diagrams for the real correction. The black dots denote the interactions between a parton and the CGC effective field.}
\end{figure}

To simplify our discussion, we will take differential cross section with both the probe parton and the fragmenting parton being a quark as an example to describe our method. The antiquark and gluon cases, which are given in Appendix \ref{sec:otherc}, can be discussed similarly. All contributions will be included in the numerical analysis. Feynman diagrams and kinematic notations are shown in Fig.~\ref{fig:diag}. We assume the proton and nucleus move in the $+z$ direction and $-z$ direction, respectively. By defining `$+$' and `$-$' components of a momentum $p$ as $p^{\pm}=\frac{1}{\sqrt{2}}(p^0\pm p^3)$, we have $p_p=(p_p^+,0,0_{\perp})$ and $p_A=(0,p_A^-,0_{\perp})$, with $p_p^+=p_A^-=\sqrt{s/2}$ and $s=(p_p+p_A)^2$ denotes the central energy of the collider. We use $k_p$, $k_h$, $p_h$ to denote momenta of the probe quark, the final state quark and the observed hadron, respectively. Based on Feynman rules in light-cone perturbation theory \cite{Bjorken:1970ah} we get the leading order differential cross section
\begin{align}\label{eq:LO}
\frac{d\sigma_{LO}}{{d^2p_{h\perp}}dy_h}=\int_{\tau}^1\frac{dz}{z^2}D_{h/q}(z)x_pf(x_p)\mathcal{F}_F(k_{h\perp};X_f),
\end{align}
with $\tau=\frac{p_h^+}{p_p^+}$, $p_h^+=z k_h^+$, $p_{h\perp}=z k_{h\perp}$, $x_p=\frac{k_p^+}{p_p^+}=\frac{\tau}{z}$, and $y_h=\frac{1}{2}\ln({p_h^+}/{p_h^-})$ denotes the rapidity of the hadron. According to these definitions, $\tau$ can be re-expressed as $\tau=\frac{p_{h\perp}}{\sqrt{s}}e^{y_h}$. $\mathcal{F}_F(k_\perp;X_f)$ is the color dipole scattering amplitude defined in the momentum space, which satisfies the momentum space BK evolution equation
\begin{align}\label{eq:BKmon}
\frac{d\mathcal{F}_F(k_{\perp};X_f)}{d\ln(1/X_f)}=\frac{\alpha_s N_c}{\pi^2} I_{\text{BK}}(k_{\perp},X_f),
\end{align}
where  $X_f$ is the CGC factorization scale and $I_{\text{BK}}(k_{\perp},X_f)$ will be given later. The momentum space BK equation is obtained from coordinate space one by doing a Fourier transformation. 

\sect{Rapidity regularization and next-to-leading order result}
\label{sec:nlo}
Although the calculation is straight forward at LO, it becomes much more complicated starting from NLO due to various divergences, including UV divergences, collinear divergences, soft divergences, and rapidity divergences. All divergences except rapidity divergences can be regularized by usual dimensional regularization.

There are many rapidity regulators in literature \cite{Collins:1981uk,Collins:1984kg,Becher:2010tm,Chiu:2011qc,Ji:2004wu,Collins:2011zzd,Becher:2011dz,Echevarria:2012pw,Li:2016axz}, but they are not convenient for us to use in the CGC factorization. We propose a new rapidity regulator as following. We shift the power of each light-cone energy denominator from $1$ to $1+\eta$, and then multiply it by a dimensional factor to compensate the mass dimension. Taking the amplitude of the diagram Fig~\ref{fig:rc.1} as an example, its light-cone energy denominator is modified as
\begin{align}\label{eq:regular}
\frac{1}{(k_h^- + k_g^- -k_p^{'-})} \to \frac{(X_f p_A^-)^{\eta}}{(k_h^- + k_g^- - k_p^{'-})^{1+\eta}},
\end{align}
where $k_h^-=\frac{k_{h\perp}^2}{2\xi k_p^+}$, $k_g^-=\frac{k_{g\perp}^2}{2(1-\xi) k_p^+}$ and $k_p^{'-}=\frac{(k_{h\perp}+k_{g\perp})^2}{2 k_p^+}$ with $\xi=k_h^+/k_p^+$. As we are interested in rapidity divergences coming from the region $\xi\to1$, we expand the small regulator $\eta$ according to $(1-\xi)^{-1+\eta}=\frac{\delta(1-\xi)}{\eta}+\frac{1}{(1-\xi)_+}+O(\eta)$, then  the right-hand side of Eq.~\eqref{eq:regular} becomes
\begin{align}\label{eq:regularexp}
\frac{1-\xi}{k_h^- + k_g^- -k_p^{'-}}\left[\frac{\delta(1-\xi)}{\eta}\left(\frac{2 X_f \frac{\tau}{z} s}{k_{g\perp}^2}\right)^\eta+\frac{1}{(1-\xi)_+}\right].
\end{align}
As the $1-\xi$ factor before the brackets will eventually cancel with other factors, the rapidity divergence appearing as $1/\eta$ is nonvanishing. 

In the calculation of differential cross sections, we perform a minimal subtraction for rapidity divergence $1/\eta$, and the subtracted rapidity divergences will be eventually absorbed by high order expansion of multipole correlators. In this way, renormalization schemes for multipole correlators are uniquely defined. After this procedure, the effect of the modification of energy denominator in Eq.~\eqref{eq:regular} is similar to introduce a cut-off $k_g^- + k_h^- -k_p^{'-}<X_f p_A^-$, which becomes $k_g^-/p_A^- <X_f $ in the rapidity divergent region $\xi\to1$. Similar effect can also be found for all other real emission diagrams, as well as loop diagrams. Therefore, with our rapidity regularization, dynamic gluons are constrained to have `$-$' momentum fraction smaller than the factorization scale $X_f$; while all other gluons, which have longer light-cone lifetime, are considered as static color sources in each collision event. This picture agrees exactly with the requirement of CGC effective theory. Therefore, the modification of light-cone energy denominators like Eq.~\eqref{eq:regular} provides a correct rapidity regularization for both real emission diagrams and loop diagrams in CGC factorization. Furthermore, like dimensional regularization, this rapidity regularization can be easily implemented in an automatic computer program. Most importantly, the regularized results are unambiguously defined, which are independent of momentum shift in loop or phase space integration. 

With dimensional regularization and rapidity regularization proposed above, we get the NLO differential cross section of the quark to quark channel,
\begin{widetext}
\begin{align}\label{eq:NLO}
\begin{split}
\frac{d\sigma_{q\to q}}{{d^2p_{h\perp}}dy_h}=&\int_{\tau}^1\frac{dz}{z^2}D_{h/q}(z)\Bigg\{x_pf(x_p)\bigg[\mathcal{F}_F(k_{h\perp};X_f)-\frac{\alpha_s N_c}{\pi^2}\ln\left(\frac{\bar{X}}{X_f}\right)I_{\text{BK}}(k_{h\perp},X_f)-\frac{\alpha_s N_c}{\pi^2}J_{\text{BK}}(k_{h\perp},X_f)\bigg]\\
				    &+\frac{\alpha_s}{2\pi^2}\frac{N_c}{2}\bigg\{\int_{\frac{\tau}{z}}^{1}d{\xi}\frac{x_p}{\xi}f(\frac{x_p}{\xi})\frac{2(1+\xi^2)}{(1-\xi)_+}I_{\text{rBK}}(k_{h\perp},X_f,\xi)+\int_{0}^{1}d{\xi}x_pf(x_p)\frac{2(1+\xi^2)}{(1-\xi)_+}I_{\text{vBK}}(k_{h\perp},X_f,\xi)\\
&+\int_{\frac{\tau}{z}}^{1}d{\xi}\frac{x_p}{\xi}f(\frac{x_p}{\xi})\pi \frac{1+\xi^2}{(1-\xi)_+}
\Big[\mathcal{F}_F(k_{h\perp};X_f)\ln\frac{k_{h\perp}^2}{\mu^2}+\frac{1}{\xi^2}\mathcal{F}_F(\frac{k_{h\perp}}{\xi};X_f)\ln\frac{k_{h\perp}^2}{\xi^2\mu^2}\Big]\\
&-\int_{0}^{1}d{\xi}x_pf(x_p)\pi \frac{2(1+\xi^2)}{(1-\xi)_+}
\mathcal{F}_F(k_{h\perp};X_f)\ln\frac{k_{h\perp}^2}{\mu^2}\\
&+\pi \int_{\frac{\tau}{z}}^{1}d\xi\frac{x_p}{\xi}f(\frac{x_p}{\xi})(1-\xi)\Big[\mathcal{F}_F(k_{h\perp};X_f)
+\frac{1}{\xi^2}\mathcal{F}_F(\frac{k_{h\perp}}{\xi};X_f)\Big]-\pi\int_{0}^{1}d\xi x_pf(x_p)2(1-\xi)\mathcal{F}_F(k_{h\perp};X_f)\bigg\}\Bigg\},
\end{split}
\end{align}
where $\bar{X}=k_{h\perp}^2/(\frac{\tau}{z}s)$, $I_{\text{BK}}(k_{\perp},X_f)=I_{\text{rBK}}(k_\perp,X_f,1)+I_{\text{vBK}}(k_\perp,X_f,1)$ with
\begin{align}\label{eq:irbk}
I_{\text{vBK}}(k_\perp,X_f,\xi)&=-\pi \int d^2l_{\perp}\mathcal{F}_F(k_{\perp};X_f)\mathcal{F}_F(l_{\perp};X_f)\ln\left[\frac{(l_{\perp}-\xi k_\perp)^2}{k_\perp^2}\right],\\
\begin{split}
I_{\text{rBK}}(k_\perp,X_f,\xi)=&\frac{1}{2}\int d^2k_{1\perp}\left[\mathcal{F}_F(k_{1\perp};X_f)\frac{1}{(\xi k_{1\perp}-k_\perp)^2}-\frac{1}{\xi^4}\mathcal{F}_F(\frac{k_\perp}{\xi};X_f)\frac{k_\perp^2}{2k_{1\perp}^2(k_{1\perp}-\frac{k_\perp}{\xi})^2}\right]\\
&+\frac{1}{2}\int d^2k_{1\perp}\left[\mathcal{F}_F(k_{1\perp};X_f)\frac{1}{(k_{1\perp}-k_\perp)^2}-\mathcal{F}_F(k_\perp;X_f)\frac{k_\perp^2}{2k_{1\perp}^2(k_{1\perp}-k_\perp)^2}\right]\\
&-\int d^2k_{1\perp}d^2l_{\perp}\mathcal{F}_F(k_{1\perp};X_f)\mathcal{F}_F(l_{\perp};X_f)
\frac{(\xi k_{1\perp}-k_\perp)\cdot(l_{\perp}-k_\perp)}{(\xi k_{1\perp}-k_\perp)^2(l_{\perp}-k_\perp)^2},	
\end{split}
\end{align}
and $J_{\text{BK}}(k_\perp,X_f)$ is defined as
\begin{align}\label{eq:jbk}
\begin{split}
J_{\text{BK}}&(k_\perp,X_f)=\int d^2k_{g\perp}\ln\left(\frac{k_{g\perp}^2}{k_{\perp}^2}\right)\left[\mathcal{F}_F(k_{g\perp}+k_{\perp};X_f)\frac{1}{k_{g\perp}^2}-\mathcal{F}_F(k_{\perp};X_f)\frac{1}{k_{g\perp}^2}\right.\\
&\left.-\int d^2l_{\perp}\mathcal{F}_F(l_{\perp}+k_{\perp};X_f)\mathcal{F}_F(k_{g\perp}+k_{\perp};X_f)\frac{k_{g\perp}\cdot l_{\perp}}{k_{g\perp}^2l_{\perp}^2}+\mathcal{F}_F(k_{\perp};X_f)\mathcal{F}_F(l_{\perp}+k_{\perp};X_f)\frac{k_{g\perp}\cdot (k_{g\perp}-l_{\perp})}{k_{g\perp}^2(k_{g\perp}-l_{\perp})^2}\right].
\end{split}
\end{align}
\end{widetext}

The last line in Eq.~\eqref{eq:NLO} is missing in all previous works. Thanks to a rigorous treatment of dimensional regularization together with $\overline{\text{MS}}$ subtraction, we find these additional terms.

The terms proportional to $I_{\text{BK}}$ or $J_{\text{BK}}$, which are direct results of our rapidity regularization, are missing in original works~\cite{Chirilli:2011km,Chirilli:2012jd}. Later on, the $J_{\text{BK}}$ term is introduced in Refs.\cite{Watanabe:2015tja,Iancu:2016vyg} by imposing so called ``kinematic constraint". In our understanding, a good factorization framework should be fully determined by regularization and subtraction schemes. The introduction of additional kinematic constraint is inconsistent with the above factorization philosophy. Practically, our method automatically introduces an approximate constraint $k_g^-<X_f p_A^-$, while the constraints introduced in Ref.\cite{Watanabe:2015tja} and Ref.\cite{Iancu:2016vyg} are effectively $k_g^-<p_A^-$ and $k_g^-<X_0 p_A^-$ with a fixed $X_0$, respectively. Due to the difference, we have an additional term $I_{\text{BK}}$, which, as we will see, enables the $X_f$ dependence in our method to be canceled out order by order in perturbation theory.

Based on the BK evolution equation Eq.~\eqref{eq:BKmon}, $X_f$ dependence of $\mathcal{F}_F(k_{h\perp};X_f)$ in the first line of Eq.~\eqref{eq:NLO} cancels with the $X_f$ dependence of the logarithm term $\ln(\bar{X}/X_f)$ in the same line, and all other $X_f$ dependences are suppressed by $O(\alpha_s^2)$. Therefore, in the NLO result, $X_f$ dependence is suppressed by one more $\alpha_s$ and thus can be safely ignored. Similar behavior is also true for higher order in $\alpha_s$ calculation. That is, by calculating to higher and higher order in $\alpha_s$, the factorization scale $X_f$ dependence becomes weaker and weaker. This is in fact the reason why high order calculation is useful in CGC factorization.

Although $X_f$ dependence is suppressed by $\alpha_s$, its choice is not unrestricted. The key for choosing $X_f$ is to avoid large logarithms at high order in $\alpha_s$ so that one has a better convergence of perturbative expansion for impact factor. For example, in the first line of Eq.~\ref{eq:NLO}, one of the NLO contribution $\alpha_s\ln(\bar{X}/X_f)$ can be significant if the difference between $\bar{X}$ and $X_f$ is large. To avoid this kind of large logarithms, $X_f$ in the first line can be chosen as $X_f=\kappa \bar{X}$ with $\kappa$ being an $O(1)$ quantity. Based on the same logic, we choose $X_f$'s from the second to the fifth lines of Eq.~\eqref{eq:NLO} to $\text{Min}\{\frac{\kappa\bar{X}}{1-\xi},X_{\text{\text{max}}}\}$ to reduce higher order corrections, where $X_{\text{\text{max}}}$ is introduced to avoid $X_f$ to be too large as $\xi\to1$.  As the $X_{\text{max}}$ dependence is at higher order in $\alpha_s$ and very tiny, which is shown in Appendix \ref{sec:irbk}, we choose $X_{\text{\text{max}}}=0.01$ in the following. In this way, the freedom to choose factorization scale $X_f$ becomes the freedom to choose the $O(1)$ quantity $\kappa$. By varying factorization scale via $\kappa$, we can estimate theoretical uncertainties of the missing higher order corrections. Note that theoretical uncertainty is hard to estimate for Refs.\cite{Watanabe:2015tja,Iancu:2016vyg} because their factorization scales are fixed. 

\sect{Numerical Results}
\label{sec:num}
For numerical calculation, we use MSTW \cite{Martin:2009iq} for proton PDF and DSS \cite{deFlorian:2007aj,deFlorian:2007ekg} for $\pi^{-}$ fragmentation functions, with collinear factorization scale chosen as $\mu=k_{h\perp}$. Dipole amplitudes $\mathcal{F}_F(k_{\perp};X_f)$ are obtained by solving the leading log BK equation with running coupling correction(i.e.\ the rcBK equation)\cite{Kovchegov:2006wf,Kovchegov:2006vj,Balitsky:2006wa} with the same parameters chosen in Ref.\cite{Fujii:2013gxa}. To make the numerical calculation reliable, we also need to modify the integrand by using the normalization condition of dipole amplitudes, detailed discussion on which is given in Appendix \ref{sec:irbk}.

\begin{figure}[]
	\begin{center}
		\includegraphics*[width=0.45\textwidth]{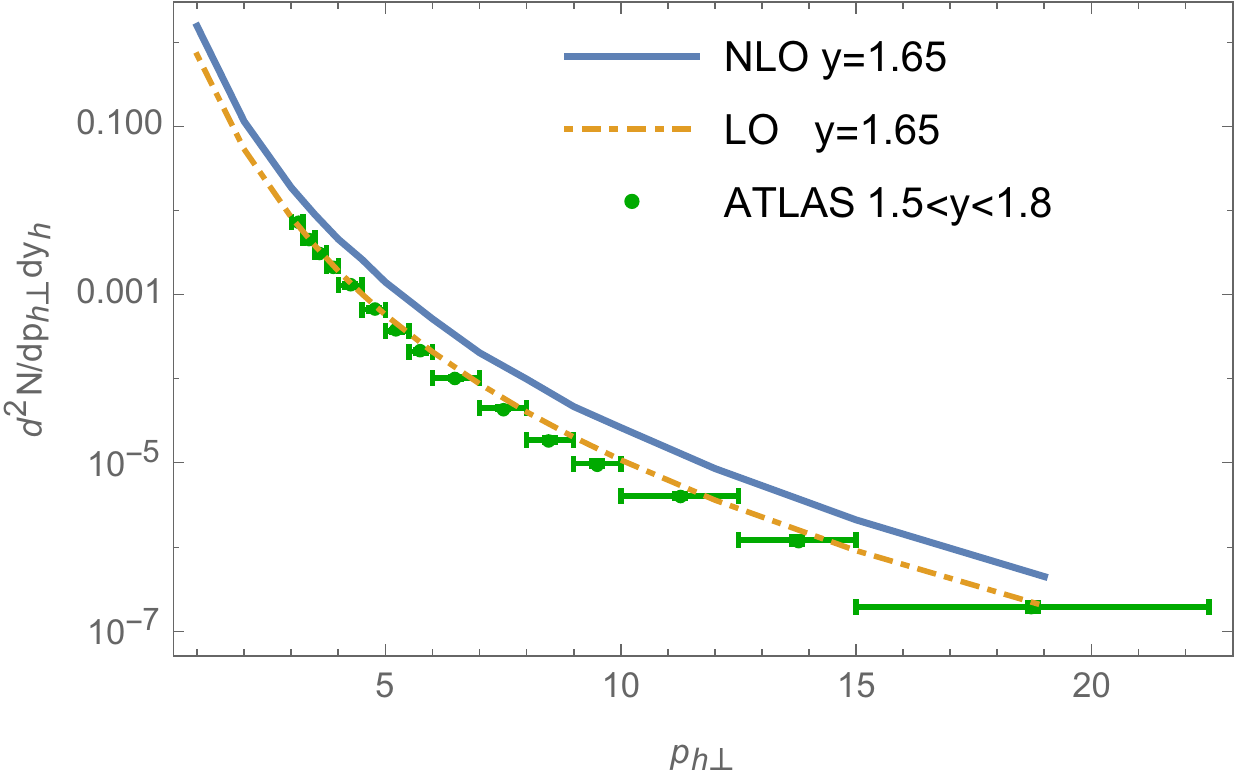}
		\caption{Comparison of the differential cross sections between the ATLAS data \cite{Aad:2016zif} and our theoretical calculation with $\sqrt{s}=5.02~\text{TeV}$, $y_h=1.65$ and $\kappa=1$.}
		\label{fig:ptdp}
	\end{center}
\end{figure}

\begin{figure}[]
	\begin{center}
		\includegraphics*[width=0.45\textwidth]{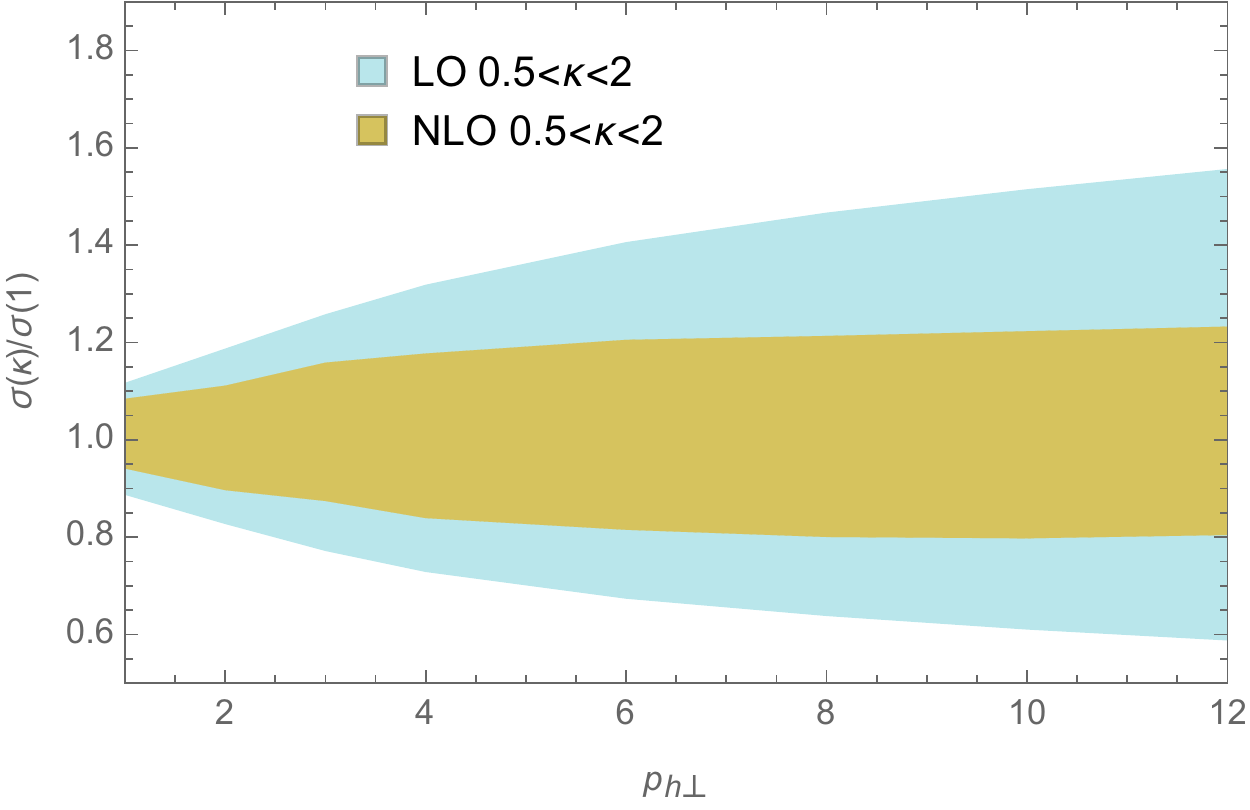}
		\caption{Predicted distributions of $d\sigma(\kappa)/d\sigma(1)$ with $\kappa$ varying from $0.5$ to $2$ for $\sqrt{s}=5.02~\text{TeV}$ and $y_h=1.65$.}
		\label{fig:cdp}
	\end{center}
\end{figure}

We take $\pi^{-}$ meson production at LHC with $\sqrt{s}=5020~\text{GeV}$ and $y_h=1.65$ as an example to show the NLO effect, which kinematic condition has also been studied in Ref.\cite{Watanabe:2015tja}.
As discussed in Ref.\cite{Watanabe:2015tja}, to compare with the forward ATLAS data \cite{Aad:2016zif}, the $\pi^{-}$ differential cross section needs to be multiplied by a pre-factor, which we choose the same as that in Ref.\cite{Watanabe:2015tja}. 
The comparison between our result with $\kappa=1$ and the forward ATLAS data \cite{Aad:2016zif} is shown in Fig.~\ref{fig:ptdp}.
It can be found that, NLO results are larger than LO. It is worth emphasizing that our results remain positive even when $p_{h\perp}$ is as large as $19~\text{GeV}$. While in Ref.\cite{Watanabe:2015tja}, by introducing kinematic constraint they can only increase positive region  from $p_{h\perp}=2.5-3$ GeV to 6 GeV. This implies that a physical choice of $X_f$, which reduces higher order contribution, 
are also crucial to solve the ``negativity" problem.

However, from the comparison with data in Fig.~\ref{fig:ptdp},  there is an illusion that the LO result is better than the NLO result. This can be understood because the initial condition of the dipole used here is obtained by fitting DIS data using LO calculation. It suggests that a global fit to determine the initial condition using complete NLO calculation is very crucial. Since formidable works are needed for this purpose, we will leave it for future study.

To estimate theoretical uncertainty introduced by CGC factorization, we vary $\kappa$ from 0.5 to 2. The ratio $d\sigma(\kappa)/d\sigma(1)$ is shown in Fig.~\ref{fig:cdp}, where $d\sigma(1)$ stands for $d^2\sigma/dp_{h\perp}^2dy_h$ with $y_h=1.65$ and $\kappa=1$. We find that the typical uncertainty at LO is 30\%-50\%, while that at NLO is less than 10\%-20\%. It tells us that NLO result indeed has much better control of theoretical uncertainties, and thus should be much more reliable.

\sect{Summary and outlook}\label{sec:sum}
In this paper, we propose a systematic method to regularize rapidity divergences in CGC factorization, which makes higher order calculation rigorous and straight forward. Due to the existence of a free factorization scale, our method can systematically estimate theoretical uncertainties caused by the missing of higher order contributions, like any standard factorization theory. As an example, we apply this method to single hadron production in pA collision. We find that the kinematics constrained term introduced by hand~\cite{Watanabe:2015tja,Iancu:2016vyg} appears automatically in our framework, but with different values. It is very crucial for the difference to enable our NLO result to have a much smaller theoretical uncertainty comparing with LO result. By choosing a physical factorization scale in the usual sense, our result also overcomes the negativity problem. 

We also find that a complete NLO fit of the initial condition of dipole amplitude is badly needed. As the method proposed in this paper can not only be used for any process but also be easily applied at the next-to-next-to-leading order and beyond, it may push the precision calculation in this field to a new stage.

\begin{acknowledgments}
	We thank C. Meng, R. Venugopalan, C.-Y. Wang, K. Watanabe, B.-W. Xiao, F. Yuan and D. Zaslavsky for many useful communications and discussions.
	The work is supported in part by
	the National Natural Science Foundation of China (Grants N. 11875071, No. 11475005 and No. 11075002),
	and the National Key Basic Research Program of China (No. 2015CB856700).
\end{acknowledgments}

\appendix

\section{Removing of $p_{h\perp}^{-2}$ terms}\label{sec:irbk}

For differential cross section $d\sigma/dp_{h\perp}^{2}$, the normalization condition
\begin{align}\label{eq:norm}
\int d^2k_{\perp}\mathcal{F}_F(k_\perp;X_f)=1\,
\end{align}
guarantees the cancellation of $p_{h\perp}^{-2}$ behavior and leaving the leading behavior at high $p_{h\perp}$ region as $p_{h\perp}^{-4}$. However in numerical calculation, the normalization condition cannot be exactly satisfied. Although the normalization condition is usually violated by a very small amount, it  can introduce significant theoretical error at high $p_{h\perp}$ region. To cure this problem, we subtract out and throw away $p_{h\perp}^{-2}$ behavior when doing numerical calculation~\footnote{We thank Bo-Wen Xiao for informing us that the same trick has been used in Ref. [38].}.

Let us first explain the existence of the $O(p_{h\perp}^{-2})$ error caused by the violation of normalization condition. For this purpose, we only need to demonstrate the existence of $O(k_{h\perp}^{-2})$ error at partonic level. Supposing $\int  d^2k_{\perp}\mathcal{F}_F(k_{\perp};X_f)=1-n(X_f)$ with nonzero $n(X_f)$ caused by numerical error or improper modeling, for sufficient large $k_{\perp}$, it is straight forward to show that $I_{\text{rBK}}(k_\perp,X_f,\xi)$ can be expanded as
\begin{align}
\frac{n(X_f)}{k_{\perp}^2}+O\left(\frac{Q^2_s(X_f)}{k_{\perp}^4}\right),
\end{align}
where we have used the fact that $\mathcal{F}_F(k_{\perp};X_f)$ is picked around $k_{\perp}^2=Q_s^2(X_f)$.
We can see that the effects of the violation will become significant if $n(X_f)\sim \frac{Q^2_s(X_f)}{k_{\perp}^2}$, which can actually happen. For example, when $\sqrt{s}=5020~\text{GeV}$, $y_h=1.65$ and $p_{h\perp}=15~\text{GeV}$, $Q_s^2/k_{h\perp}^2=(z^2Q_s^2)/p_{h\perp}^2$ can be as small as $10^{-4}$ in some regions of $z$; at the same time, for some values of $X_f$, like $0.015$, $n(X_f)$ can reach $0.0778$. Then we have $n(X_f)\gg \frac{Q^2_s(X_f)}{k_{\perp}^2}$. Although this condition is true only in a small region, the net effect may still be significant. 

To deal with the above problem, we need a better form of $I_{\text{rBK}}(k_\perp,X_f,\xi)$ in numerical implementation. One of the possible choices is
\begin{widetext}\begin{align}\label{eq:irbkq}
	\begin{split}
	I^Q_{\text{rBK}}(k_\perp,X_f,\xi,Q)=&\frac{1}{2}\int d^2k_{1\perp}[\mathcal{F}_F(k_{1\perp};X_f)\frac{1}{(\xi k_{1\perp}-k_\perp)^2}-\mathcal{F}_F(k_{1\perp};X_f)\frac{1}{k_{\perp}^2+Q^2}-\frac{1}{\xi^4}\mathcal{F}_F(\frac{k_\perp}{\xi};X_f)\frac{k_\perp^2}{2k_{1\perp}^2(k_{1\perp}-\frac{k_\perp}{\xi})^2}]\\
	&+\frac{1}{2}\int d^2k_{1\perp}[\mathcal{F}_F(k_{1\perp};X_f)\frac{1}{(k_{1\perp}-k_\perp)^2}-\mathcal{F}_F(k_{1\perp};X_f)\frac{1}{k_{\perp}^2+Q^2}-\mathcal{F}_F(k_\perp;X_f)\frac{k_\perp^2}{2k_{1\perp}^2(k_{1\perp}-k_\perp)^2}]\\&-\int d^2k_{1\perp}d^2l_{\perp}\mathcal{F}_F(k_{1\perp};X_f)\mathcal{F}_F(l_{\perp};X_f)
	[\frac{(\xi k_{1\perp}-k_\perp)\cdot(l_{\perp}-k_\perp)}{(\xi k_{1\perp}-k_\perp)^2(l_{\perp}-k_\perp)^2}-\frac{1}{k_{\perp}^2+Q^2}],	
	\end{split}
	\end{align}\end{widetext}
where $Q$ is an $O(Q_s)$ value. $I^Q_{\text{rBK}}$ contains no $O(k_{\perp}^{-2})$ violation terms and is equivalence to $I_{\text{rBK}}$ if the normalization is not violated.

Since the normalization violation does exist in numerical calculation, it is necessary to study the dependence on the additional parameter $Q$. We use $\sigma_1(X_{\text{max}},Q)$ to denote the result of the term proportional to $I^Q_{\text{rBK}}$ with $\sqrt{s}=5020~\text{GeV}$, $y_h=1.65$, $p_{h\perp}=15~\text{GeV}$ and $\kappa=1$. $\sigma_1(X_{\text{max}},Q)/\sigma_1(0.01,2)$ under different $X_{\text{max}}$ and $Q$ choices are shown in Table.~\ref{tal:dponc}. It shows that the dependence on $X_{\text{max}}$ and $Q$ is very small when $Q\sim Q_s$. We thus choose $Q=2~\text{GeV}$ and $X_{\text{max}}=0.01$ in the numerical calculation.

\begin{table}[]
	\caption{The dependence of $\sigma_1(X_{\text{max}},Q)/\sigma_1(0.01,2)$ on $Q$ and $X_{\text{max}}$}
	\begin{tabular}[c]{@{}|c|c|c|c|c|}
		\toprule
		\diagbox{$Q$ (GeV)}{$X_{\text{max}}$}   &$0.01$ & $0.02$ & $0.1$ & $1$\\
		\colrule
		$2$  &1 &0.9991 &0.9914 &0.9851\\ \colrule
		$3$ &1.005 &1.024 &1.030 &1.009\\ \colrule
		$4$ &1.003 &1.057 &1.081 &1.065\\
		\botrule
	\end{tabular}
	\label{tal:dponc}
\end{table}

\begin{widetext}
	\section{Perturbative calculation of other channels}\label{sec:otherc}
	In this appendix, we list our impact factors of $g\to g$, $q\to g$ and $g\to q$ channels up to $O(\alpha_s)$ precision, where the former parton comes from the proton and the later one is the fragmenting parton. 
	
	\subsection{Result of $g\to g$ channel}
	The NLO result of the $g\to g$ channel is 
	\begin{align}\label{eq:finalgg}
	\begin{split}
	\frac{d\sigma_{g\to g}}{{d^2p_{h\perp}}dy_h}=&\int_{\tau}^1\frac{dz}{z^2}D_{h/g}(z)\{x_pG(x_p)[\mathcal{F}_A(k_{h\perp};X_f)-\frac{2\alpha_s N_c}{\pi^2}\ln\left(\frac{\bar{X}}{X_f}\right)I_{ggBK}(k_{h\perp},X_f)-\frac{2\alpha_s N_c}{\pi^2}J_{ggBK}(k_{h\perp},X_f)]\\
	&+\frac{\alpha_sN_c}{\pi^2}\{\int_{\frac{\tau}{z}}^{1}d{\xi}\frac{x_p}{\xi}G(\frac{x_p}{\xi})2[\frac{\xi}{(1-\xi)^+}+\frac{1-\xi}{\xi}+\xi(1-\xi)][I_{rggBK}(k_{h\perp},X_f,\xi)]\\
	&+2\pi\int_{0}^{1}d{\xi}x_pG(x_p)[\frac{\xi}{(1-\xi)^+}+\frac{1}{2}\xi(1-\xi)]I_{vggBK}(k_{h\perp},X_f,\xi)\\
	&+\int_{\frac{\tau}{z}}^{1}d{\xi}\frac{x_p}{\xi}G(\frac{x_p}{\xi}) \pi[\frac{\xi}{(1-\xi)^+}+\frac{1-\xi}{\xi}+\xi(1-\xi)]
	[\mathcal{F}_A(k_{h\perp};X_f)\ln\frac{k_{h\perp}^2}{\mu^2}+\frac{1}{\xi^2}\mathcal{F}_A(\frac{k_{h\perp}}{\xi};X_f)\ln\frac{k_{h\perp}^2}{\xi^2\mu^2}]\\
	&-2\pi\int_{0}^{1}d{\xi}x_pG(x_p) \{[\frac{\xi}{(1-\xi)^+}+\frac{1}{2}\xi(1-\xi)]+\frac{N_fT_R}{2N_c}[\xi^2+(1-\xi)^2]\}
	\mathcal{F}_A(k_{h\perp};X_f)\ln\frac{k_{h\perp}^2}{\mu^2}\}\\
	&-\frac{\alpha_s}{\pi}N_f T_R \int_{0}^{1}d{\xi}x_pG(x_p)\{[\xi^2+(1-\xi)^2]\int d^2l_{\perp}\mathcal{F}_F(k_{h\perp}-l_{\perp};X_f)\mathcal{F}_F(l_{\perp};X_f)\ln[\frac{(l_{\perp}-\xi k_{h\perp})^2}{k_{h\perp}^2}]\\
	&+2(\xi-\xi^2)\mathcal{F}_A(k_{h\perp};X_f)\}\},
	\end{split}
	\end{align}
	where $\mathcal{F}_A(k_\perp;X_f)$ denotes the dipole amplitude in adjoint representation, which can be related to dipole amplitude in fundamental representation by
	\begin{align}
	\mathcal{F}_A(k_\perp;X_f)=\int d^{D-2} k_{1\perp}\mathcal{F}_F(k_\perp;X_f)\mathcal{F}_F(k_\perp-k_{1\perp};X_f).
	\end{align}
	In Eq.~\eqref{eq:finalgg} we define
	\begin{align}
	I_{ggBK}(k_\perp,X_f)=I_{rggBK}(k_\perp,X_f,1)+I_{vggBK}(k_\perp,X_f,1),
	\end{align}
	with
	\begin{align}
	\begin{split}
	I_{rggBK}(k_\perp,X_f,\xi)=&\frac{1}{2}\int d^2k_{1\perp}[\mathcal{F}_A(k_{1\perp};X_f)\frac{1}{(\xi k_{1\perp}-k_\perp)^2}-\frac{1}{\xi^4}\mathcal{F}_A(\frac{k_\perp}{\xi};X_f)\frac{k_\perp^2}{2k_{1\perp}^2(k_{1\perp}-\frac{k_\perp}{\xi})^2}]\\
	&+\frac{1}{2}\int d^2k_{1\perp}[\mathcal{F}_A(k_{1\perp};X_f)\frac{1}{(k_{1\perp}-k_\perp)^2}-\mathcal{F}_A(k_\perp;X_f)\frac{k_\perp^2}{2k_{1\perp}^2(k_{1\perp}-k_\perp)^2}]\\
	&-\int d^2k_{1\perp}d^2l_{\perp}d^2l_{1\perp}\mathcal{F}_F(l_{1\perp};X_f)\mathcal{F}_F(k_{1\perp}+l_{1\perp};X_f)\mathcal{F}_F(l_{\perp}+l_{1\perp};X)
	\frac{(\xi k_{1\perp}-k_\perp)\cdot(l_{\perp}-k_\perp)}{(\xi k_{1\perp}-k_\perp)^2(l_{\perp}-k_\perp)^2}],
	\end{split}
	\end{align}
	\begin{align}
	I_{vggBK}(k_\perp,X_f,\xi)=-\int d^2l_{\perp}d^2l_{1\perp}\mathcal{F}_F(k_{\perp}+l_{1\perp};X_f)\mathcal{F}_F(l_{\perp}+l_{1\perp};X_f)\mathcal{F}_F(l_{1\perp};X_f)\ln[\frac{(l_{\perp}-\xi k_\perp)^2}{k_\perp^2}],
	\end{align}
	and
	\begin{align}\label{eq:jggbk}
	\begin{split}
	J_{ggBK}(k_\perp,X_f)=&\int d^2k_{g\perp}\int d^2l_{1\perp}\mathcal{F}_F(l_{1\perp};X_f)\{[\mathcal{F}_F(k_{g\perp}+k_{\perp}+l_{1\perp};X_f)-\mathcal{F}_F(k_{\perp}+l_{1\perp};X_f)]\frac{1}{k_{g\perp}^2}\ln\left(\frac{k_{g\perp}^2}{k_{\perp}^2}\right)\\
	&-\int d^2l_{\perp}\mathcal{F}_F(l_{\perp}+k_{\perp}+l_{1\perp};X_f)\mathcal{F}_F(k_{g\perp}+k_{\perp}+l_{1\perp};X_f)\frac{k_{g\perp}\cdot l_{\perp}}{k_{g\perp}^2l_{\perp}^2}\ln\left(\frac{k_{g\perp}^2}{k_{\perp}^2}\right)\\
	&+\int d^2l_{\perp}\mathcal{F}_F(k_{\perp}+l_{1\perp};X_f)\mathcal{F}_F(l_{\perp}+k_{\perp}+l_{1\perp};X_f)\frac{k_{g\perp}\cdot (k_{g\perp}-l_{\perp})}{k_{g\perp}^2(k_{g\perp}-l_{\perp})^2}\ln\left(\frac{k_{g\perp}^2}{k_{\perp}^2}\right)\}.
	\end{split}
	\end{align}
	
	\subsection{Result of $q\to g$ channel}
	The $q\to g$ channel gives
	\begin{align}\label{eq:finalqg}
	\begin{split}
	\frac{d\sigma_{q\to g}}{{d^2p_{h\perp}}dy_h}=&\frac{\alpha_sN_c}{4\pi^2}\int_{\tau}^1\frac{dz}{z^2}D_{h/g}(z)\int_{\frac{\tau}{z}}^{1}d{\xi}\frac{x_p}{\xi}f(\frac{x_p}{\xi})\{\mathcal{P}_{gq}(\xi)[\pi \mathcal{F}_A(k_{h\perp};X_f)\ln\frac{k_{h\perp}^2}{\mu^2}+\frac{\pi}{\xi^2}\mathcal{F}_F(\frac{k_{h\perp}}{\xi};X_f)\ln\frac{k_{h\perp}^2}{\xi^2\mu^2}]\\
	&+\mathcal{P}_{gq}(\xi)\int d^2k_{1\perp}[\mathcal{F}_A(k_{1\perp};X_f)\frac{1}{(k_{1\perp}-k_{h\perp})^2}-\mathcal{F}_A(k_{h\perp};X_f)\frac{k_{h\perp}^2}{2k_{1\perp}^2(k_{1\perp}-k_{h\perp})^2}]\\
	&+\mathcal{P}_{gq}(\xi)\int d^2k_{1\perp}[\mathcal{F}_F(k_{1\perp};X_f)\frac{1}{(\xi k_{1\perp}-k_{h\perp})^2}-\frac{1}{\xi^4}\mathcal{F}_F(\frac{k_{h\perp}}{\xi};X_f)\frac{k_{h\perp}^2}{2k_{1\perp}^2(k_{1\perp}-\frac{k_{h\perp}}{\xi})^2}]\\
	&-2\mathcal{P}_{gq}(\xi)\int d^2k_{1\perp}d^2l_{\perp}\mathcal{F}_F(k_{1\perp};X_f)\mathcal{F}_F(k_{1\perp}-l_{\perp};X_f)
	\frac{(\xi k_{1\perp}-k_{h\perp})\cdot(l_{\perp}-k_{h\perp})}{(\xi k_{1\perp}-k_{h\perp})^2(l_{\perp}-k_{h\perp})^2}\\
	&+\pi\xi[\mathcal{F}_A(k_{h\perp};X_f)+\frac{1}{\xi^2}\mathcal{F}_F(\frac{k_{h\perp}}{\xi};X_f)]\},
	\end{split}
	\end{align}
	where $\mathcal{P}_{gq}(\xi)=\frac{1}{\xi}[1+(1-\xi)^2]$.
	
	\subsection{Result of $g\to q$ channel}
	The $g\to q$ channel gives 
	\begin{align}\label{eq:finalgq}
	\begin{split}
	\frac{d\sigma_{g\to q}}{{d^2p_{h\perp}}dy_h}=&\frac{\alpha_sT_R}{2\pi^2}\int_{\tau}^1\frac{dz}{z^2}D_{h/q}(z)\int_{\frac{\tau}{z}}^{1}d{\xi}\frac{x_p}{\xi}G(\frac{x_p}{\xi})\{\mathcal{P}_{qg}(\xi)[\pi \mathcal{F}_F(k_{h\perp};X_f)\ln\frac{k_{h\perp}^2}{\mu^2}+\frac{\pi}{\xi^2}\mathcal{F}_A(\frac{k_{h\perp}}{\xi};X_f)\ln\frac{k_{h\perp}^2}{\xi^2\mu^2}]\\
	&+\mathcal{P}_{qg}(\xi)\int d^2k_{1\perp}[\mathcal{F}_F(k_{1\perp};X_f)\frac{1}{(k_{1\perp}-k_{h\perp})^2}-\mathcal{F}_F(k_{h\perp};X_f)\frac{k_{h\perp}^2}{2k_{1\perp}^2(k_{1\perp}-k_{h\perp})^2}]\\
	&+\mathcal{P}_{qg}(\xi)\int d^2k_{1\perp}[\mathcal{F}_A(k_{1\perp};X_f)\frac{1}{(\xi k_{1\perp}-k_{h\perp})^2}-\frac{1}{\xi^4}\mathcal{F}_A(\frac{k_{h\perp}}{\xi};X_f)\frac{k_{h\perp}^2}{2k_{1\perp}^2(k_{1\perp}-\frac{k_{h\perp}}{\xi})^2}]\\
	&-2\mathcal{P}_{qg}(\xi)\int d^2k_{1\perp}d^2l_{\perp}\mathcal{F}_F(l_{\perp};X_f)\mathcal{F}_F(k_{1\perp}-l_{\perp};X_f)
	\frac{(\xi k_{1\perp}-k_{h\perp})\cdot(l_{\perp}-k_{h\perp})}{(\xi k_{1\perp}-k_{h\perp})^2(l_{\perp}-k_{h\perp})^2}\\
	&-\pi({2\xi^2-2\xi})[\mathcal{F}_F(k_{h\perp};X_f)+\frac{1}{\xi^2}\mathcal{F}_A(\frac{k_{h\perp}}{\xi};X_f)]\},
	\end{split}
	\end{align}
	where $\mathcal{P}_{qg}(\xi)=[\xi^2+(1-\xi)^2]$.
\end{widetext}



\providecommand{\href}[2]{#2}\begingroup\raggedright\endgroup

\end{document}